\documentclass[pra,twocolumn,shownopacs,amssymb]{revtex4}
\usepackage{graphicx}
\usepackage{amsmath}

\begin{document}

\title{Origin of flat-band superfluidity on the Mielke checkerboard lattice}

\author{M. Iskin}
\affiliation{Department of Physics, Ko\c{c} University, Rumelifeneri Yolu, 
34450 Sar\i yer, Istanbul, Turkey}

\date{\today}

\begin{abstract}

The Mielke checkerboard is known to be one of the simplest two-band 
lattice models exhibiting an energetically flat band that is in touch with a 
quadratically dispersive band in the reciprocal space, i.e., its flat band 
is not isolated. Motivated by the growing interest in understanding the 
origins of flat-band superfluidity in various contexts, here we provide 
an in-depth analysis showing how the mean-field BCS correlations prevail 
in this particular model. Our work reveals the quantum-geometric origin 
of flat-band superfluidity through uncovering the leading role by 
a band-structure invariant, i.e., the so-called quantum metric tensor 
of the single-particle bands, in the inverse effective mass tensor of the 
Cooper pairs. 

\end{abstract}

\maketitle

\section{Introduction}
\label{sec:intro}

Given the successful realizations of Kagome~\cite{jo12, nakata12, li18}
and Lieb~\cite{taie15, diebel16, kajiwara16, ozawa17} lattices in a 
variety of settings that include optical potentials and photonic wave guides, 
flat-band physics is gradually becoming one of the central themes in 
modern physics~\cite{liu14, leykam18}. From the many-body physics 
perspective, there has been a growing interest in understanding its 
ferromagnetism~\cite{tasaki98}, fractional quantum Hall
physics~\cite{parameswaran13}, and superconductivity~\cite{khodel90, 
kopnin11, iglovikov14, tovmasyan18, mondaini18}. In addition, the 
electronic band structure of `twisted bilayer graphene' also exhibits flat 
bands near zero Fermi energy~\cite{cao18, yankowitz19}, and these
flat regions are believed to characterize most of the physical properties 
of this two-dimensional wonder material, e.g., its relatively high 
superconducting transition temperature. In fact, the theoretical interest 
on flat-band superconductors goes back a long time since they 
were predicted to be a plausible route for our ultimate goal of reaching 
a room-temperature superconductor~\cite{khodel90, kopnin11}. This 
expectation was based on the naive BCS theory, implied simply by the 
relatively high single-particle density of states for narrower bands. 

It is important to emphasize that not only the physical mechanism for the 
origin of flat-band superfluidity was missing in these earlier works but also 
it was not absolutely clear whether superfluidity can exist in a flat band.
This is because superfluidity is strictly forbidden if the allowed single-particle
states in the reciprocal ($\mathbf{k}$) space are only from a single flat band.
These issues were partially resolved back in 2015 once the superfluid 
(SF) density was shown to depend not only on the energy dispersion but 
also on the Bloch wave functions of a lattice Hamiltonian~\cite{torma15}. 
The latter dependence is a direct result of the band geometry in such 
a way that the superfluidity may succeed in an isolated flat band 
only in the presence of other bands, through the interband 
processes~\cite{torma15, torma16, torma17a, torma19}. These 
geometric effects are characterized by a band-structure invariant known 
as the quantum metric tensor~\cite{provost80, berry89, thouless98}.

Following up this line of research in various directions~\cite{torma15, 
torma16, torma17a, torma19, iskin17, iskin18b, iskin19}, here we expose 
the quantum-geometric origin of flat-band superfluidity in the Mielke 
checkerboard lattice~\cite{mielke91, montambaux18}, which is known 
to be one of the simplest two-band lattice models exhibiting an 
energetically flat band that is in touch with a quadratically dispersive 
one in $\mathbf{k}$ space.
For instance, in the weak-binding regime of arbitrarily low interactions, 
we show that the inverse effective mass tensor of the Cooper pairs 
is determined entirely by a $\mathbf{k}$-space sum over the quantum 
metric tensor, but with a caveat for the non-isolated flat band. 
Since the effective band mass of a non-interacting 
particle is infinite in a flat band, our finding illuminates 
the physical mechanism behind how the mass of the SF carriers becomes 
finite with a finite interaction. That is how the quantum geometry is 
responsible for it through the interband processes. When the interaction 
increases, we also show that the geometric interband contribution 
gradually gives way to the conventional intraband one, eventually 
playing an equally important role in the strong-binding regime. 
Given that the pair mass plays direct roles in a variety of SF properties 
that are thoroughly discussed in this paper but not limited to them, 
our work also illuminates how the mean-field BCS correlations prevail 
in a non-isolated flat band. 

The rest of the paper is organized as follows. Starting with a brief 
introduction to the checkerboard lattice of interest in Sec.~\ref{sec:tb}, 
we outline the single-particle problem and the mean-field Hamiltonian 
in Sec.~\ref{sec:mfh}, overview the resultant self-consistency equations 
in Sec.~\ref{sec:sce}, and then comment on the strong-binding regime of 
molecular pairs in Sec.~\ref{sec:sbmp}. Our numerical calculations are 
presented in Secs.~\ref{sec:na} and~\ref{sec:fbs}, and they are supported 
by an in-depth analysis showing how the flat-band superfluidity prevails 
both in the Mielke checkerboard lattice in Sec.~\ref{sec:mclm} and in 
its low-energy continuum approximation in Sec.~\ref{sec:elcm}. The paper 
ends with a summary of our conclusions in Sec.~\ref{sec:conc}.

\section{Theoretical Background}
\label{sec:tb}

\begin{figure}[htbp]
\includegraphics[scale=0.29]{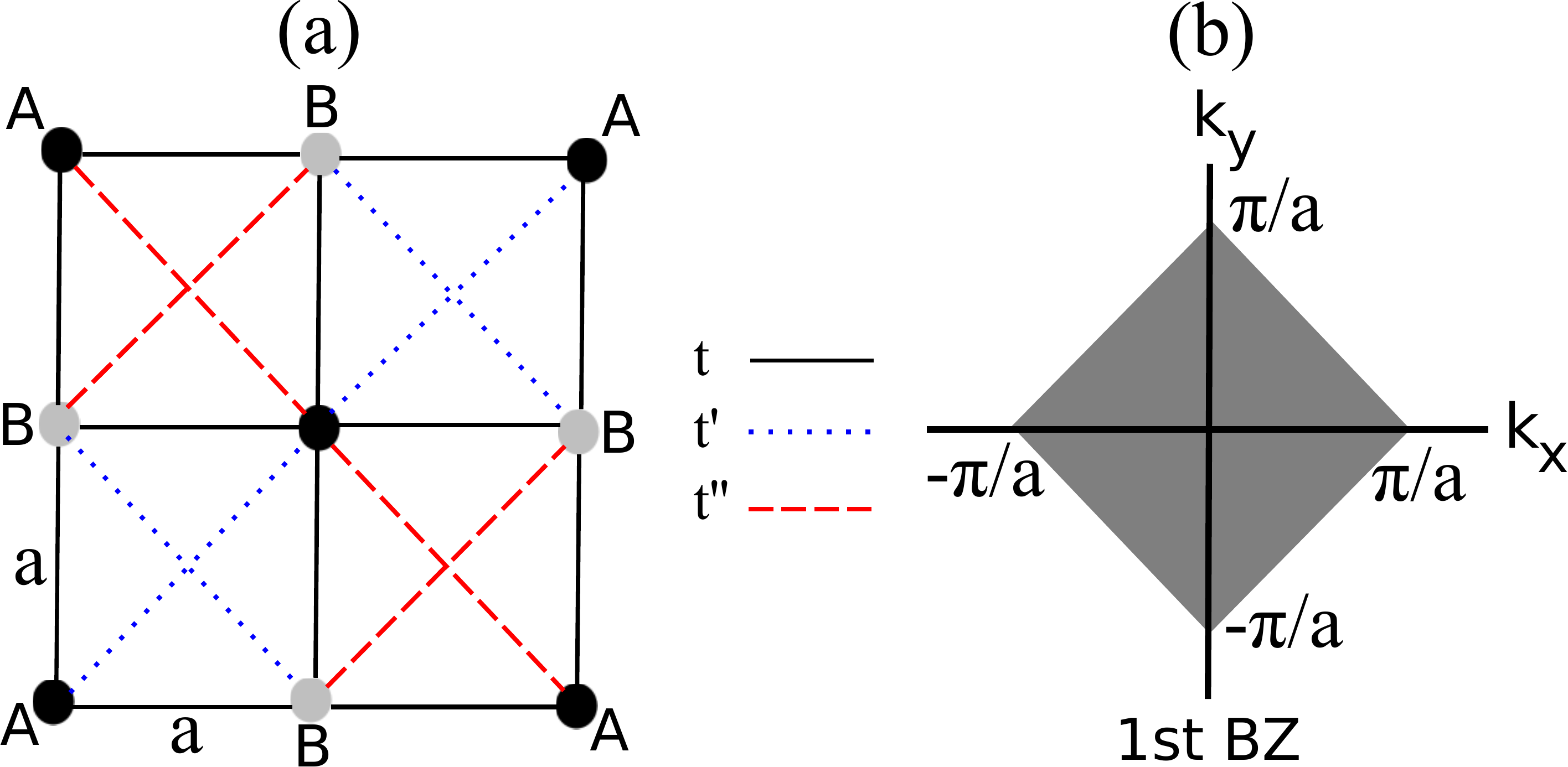}
\caption{(color online)
\label{fig:cb-lattice}
Sketches of (a) the crystal structure together with the hopping parameters
in the real space, and (b) the corresponding first Brillouin zone in the 
reciprocal space.
}
\end{figure}

The Mielke checkerboard lattice belongs to a special class of lattices known 
as line graphs, and they are by construction in such a way that a destructive
interference between nearest-neighbor sites gives rise to a flat 
band~\cite{mielke91, montambaux18}. For instance, the line graph of a 
given lattice is obtained as follows: first introduce new lattice points in the 
middle of every bond in the original lattice, and then connect only those 
new points that belong to the nearest-neighbor bonds in the original lattice, 
i.e., the ones sharing a common site. This is such that the line graph of a 
honeycomb lattice is a Kagome lattice, and that of the checkerboard lattice 
is a Mielke checkerboard~\cite{liu14, leykam18}. 
In this paper, we first consider a more general lattice that is sketched in
Fig.~\ref{fig:cb-lattice}(a), and study the Mielke checkerboard case as 
one of its limits. 

To describe the hopping kinematics of the single-particle problem on such 
a lattice, we choose the primitive lattice vectors
$
\mathbf{a_1} = (a, -a)
$
and
$
\mathbf{a_2} = (a, a)
$
together with a basis of two sites. We are particularly interested in the 
competition between the following hopping parameters: if the location 
of a given site on the $A$ sublattice is $(0, 0)$ then the intersublattice 
hopping $t$ connects it to the nearest-neighbor sites 
$\{(\pm a, 0), (0, \pm a)\}$ on the $B$ sublattice, and the intrasublattice 
ones $t'$ and $t''$ connect it to the next-nearest-neighbor sites 
$\{(a, a), (-a, -a)\}$ and $\{(a, -a), (-a, a)\}$, respectively, on the $A$ sublattice.
In this paper, we assume $t > 0$ and $t' > t'' \ge 0$ without loss of generality.
The corresponding reciprocal lattice vectors
$
\mathbf{b_1} = (\pi/a, -\pi/a) 
$
and
$
\mathbf{b_2} = (\pi/a, \pi/a)
$
give rise to the first Brillouin zone that is sketched in Fig.~\ref{fig:cb-lattice}(b).

Depending on the interplay of these hopping parameters, such a crystal 
structure is known to exhibit two single-particle energy bands with very 
special features, e.g., an energetically flat band that shares quadratic 
touching points with a dispersive band in the reciprocal space, 
as discussed next.

\subsection{Mean-field Hamiltonian}
\label{sec:mfh}

In the reciprocal space, the hopping contribution to the Hamiltonian can be 
written as
$
H_0 = \sum_\mathbf{\sigma k} \psi_{\sigma \mathbf{k}}^\dagger h_\mathbf{k} \psi_{\sigma \mathbf{k}},
$
where 
$
\psi_{\sigma \mathbf{k}}^\dagger = (c_{\sigma A \mathbf{k}}^\dagger \, c_{\sigma B \mathbf{k}}^\dagger)
$
is the creation operator for a two-component sublattice spinor, and
\begin{align}
\label{eqn:hk}
h_\mathbf{k} = d_\mathbf{k}^0 \tau_0 + \mathbf{d_\mathbf{k}} \cdot \boldsymbol{\tau}
\end{align}
is the Hamiltonian density with $\tau_0$ a unit matrix, 
$
\boldsymbol{\tau} = (\tau_x, \tau_z)
$ 
a vector of Pauli matrices, and
$
\mathbf{d_k} = (d_\mathbf{k}^x, d_\mathbf{k}^z).
$ 
While the diagonal elements of $h_\mathbf{k}$ are due solely to the 
next-nearest-neighbor hoppings, the off-diagonal elements are due 
to the nearest-neighbor ones in such a way that
\begin{align}
\label{eqn:dk0}
d_\mathbf{k}^0 &= - 2(t'+t'') \cos(k_x a) \cos(k_y a), \\
\label{eqn:dkx}
d_\mathbf{k}^x &= - 2t \cos(k_x a) - 2t \cos(k_y a), \\
\label{eqn:dkz}
d_\mathbf{k}^z &= 2(t'-t'') \sin(k_x a) \sin(k_y a).
\end{align}
The single-particle energy bands are determined by 
$
\varepsilon_{s \mathbf{k}} = d_\mathbf{k}^0 + s d_\mathbf{k},
$
where $s = \pm$ denotes the upper/lower band and $d_\mathbf{k} = |\mathbf{d_k}|$. 
These bands touch each other at the following points 
$
\mathbf{k} \equiv \{(\pm \pi/a, 0), (0, \pm \pi/a)\}
$
in the reciprocal space, i.e., at the four corners of the first Brillouin zone, 
where $d_\mathbf{k} = 0$ and $\varepsilon_{s \mathbf{k}} = 2(t'+t'')$. 
In particular when $t' = t > t'' = 0$, the band structure reduces to 
$
\varepsilon_{+, \mathbf{k}} = 2t
$
and
$
\varepsilon_{-, \mathbf{k}} = -2t - 4t \cos(k_xa) \cos(k_ya),
$
and therefore, a flat upper band that shares quadratic touching points 
with a dispersive lower band emerges. This is known to be one of the 
simplest two-band lattice models exhibiting a flat band.

\begin{figure*}[htbp]
\includegraphics[scale=0.12]{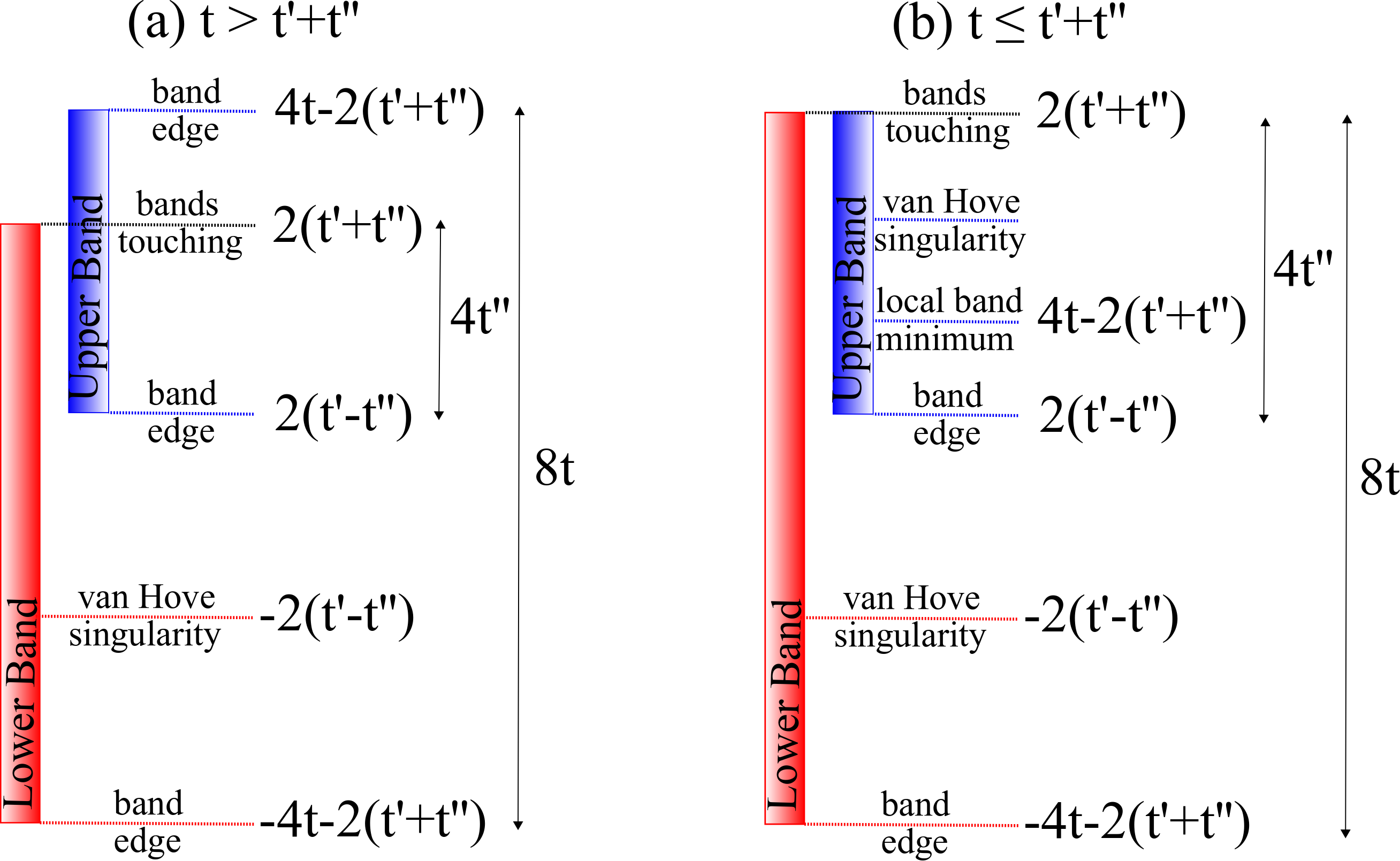}
\caption{(color online)
\label{fig:bandsketch}
Sketches of the single-particle bands depending on whether the hopping
parameters are related to each other through $t > t'+t''$ or not. Here, 
and throughout this paper, we assume $t' > t'' \ge 0$ without loss of generality.
}
\end{figure*}

For a given relation between the hopping parameters, the relative 
positions of these bands together with some of their important features 
are sketched in Fig.~\ref{fig:bandsketch}. For instance, when $t > t'+t''$, 
sketch~\ref{fig:bandsketch}(a) implies that the upper band edges of 
the lower band touch to the local band minima of the upper band at the 
energy $2(t'+t'')$. The band width $4(t-t')$ of the upper band is narrow 
compared to that $4(t+t'+t'')$ of the lower band.
Thus, when $t > t' > t'' = 0$, the upper band edges of the lower 
band touch precisely to the lower band edges of the upper band at
the energy $2t'$.
On the other hand, when $t \le t'+t''$, sketch~\ref{fig:bandsketch}(b) 
implies that the upper band edges of the lower band touch to the upper 
band edges of the upper band at the energy $2(t'+t'')$. 
Since the band width $4t''$ of the upper band is controlled 
independently of that $8t$ of the lower band, an energetically 
quasi-flat upper band eventually appears in the $t''/t \to 0$ limit.

In direct analogy with the recent 
works~\cite{torma17a, iskin17, iskin18b, iskin19}, the $\mathbf{k}$-space 
structure of $\mathbf{d_k}$ suggests that the quantum geometry of the reciprocal 
space is nontrivial as long as $t' \ne t''$ for a given $t \ne 0$, i.e., the 
quantum metric tensor plays an important role in the formation of Cooper 
pairs and the related phenomena. To illustrate the quantum metric effects on 
the many-body problem, here we consider the case of attractive onsite 
interactions between $\uparrow$ and $\downarrow$ particles, and take
them into account within the BCS mean-field approximation for pairing. 
The resultant Hamiltonian for the stationary Cooper pairs 
can be written as
\begin{align}
H_\textrm{mf} &= \sum_\mathbf{k} \Psi_\mathbf{k}^\dagger 
\left(
\begin{array}{cc}
  h_\mathbf{k} - \mu \tau_0& \boldsymbol{\Delta} \\
  \boldsymbol{\Delta}^\dagger  & -h_{-\mathbf{k}}^* + \mu \tau_0 \\
\end{array}
\right)
\Psi_\mathbf{k} \nonumber \\
& + 2 \sum_\mathbf{k} (d_0^{\mathbf{k}} - \mu) + M \frac{\Delta^2}{U},
\label{eqn:ham}
\end{align}
where
$
\Psi_\mathbf{k}^\dagger = (\psi_{\uparrow \mathbf{k}}^\dagger \, \psi_{\downarrow, -\mathbf{k}})
$
is a four-component spinor operator, 
$
\boldsymbol{\Delta}_{S S'} = \Delta_S \delta_{S S'}
$
is the order parameter with $\delta_{ij}$ the Kronecker-delta, $\mu$ is the 
chemical potential, $M$ is the number of lattice sites, and $U \ge 0$
is the strength of the interaction. It turns out that 
$
\Delta_S = (2U/M) \sum_{\mathbf{k}} 
\langle c_{\downarrow S,-\mathbf{k}} c_{\uparrow S \mathbf{k}} \rangle
$
with $\langle \ldots \rangle$ denoting the thermal average is uniform for the 
entire lattice, and we take $\Delta = \Delta_S$ as real without loss of generality.

\subsection{Self-consistency equations}
\label{sec:sce}

Given the mean-field Hamiltonian above by Eq.~(\ref{eqn:ham}), a compact 
way to express the resultant self-consistency (order parameter and number) 
equations are~\cite{iskin19},
\begin{align}
\label{eqn:op}
1 &= \frac{U}{2M} \sum_{s \mathbf{k}} \frac{\mathcal{X}_{s \mathbf{k}}}{E_{s \mathbf{k}}}, \\
F &= 1 - \frac{1}{M} \sum_{s \mathbf{k}} 
\frac{\mathcal{X}_{s \mathbf{k}}}{E_{s\mathbf{k}}} \xi_{s \mathbf{k}},
\label{eqn:filling}
\end{align}
where
$
\xi_{s \mathbf{k}}=\varepsilon_{s \mathbf{k}} - \mu
$
is the shifted dispersion,
$
\mathcal{X}_{s \mathbf{k}} = \tanh [E_{s \mathbf{k}}/(2T)]
$
is a thermal factor with $k_\textrm{B} \to 1$ the Boltzmann constant and $T$ the 
temperature,
$
E_{s \mathbf{k}} = \sqrt{\xi_{s \mathbf{k}}^2 + \Delta^2}
$
is the quasi-particle energy spectrum, and $0 \le F = N/M \le 2$ is the total particle 
filling with 
$
N = \sum_{\sigma S \mathbf{k}}
\langle c_{\sigma S \mathbf{k}}^\dag c_{\sigma S \mathbf{k}} \rangle
$
the total number of particles.
Thus, the naive BCS mean-field theory corresponds to the self-consistent solutions 
of Eqs.~(\ref{eqn:op}) and~(\ref{eqn:filling}) for $\Delta$ and $\mu$ for any given 
set of model parameters $t'$, $t''$, $U$, $F$, and $T$. For instance, the critical 
BCS transition temperature $T_\textrm{BCS}$ is determined by the condition 
$\Delta \to 0^+$, and it gives a reliable estimate for the critical SF transition 
temperature in the weak-binding regime where $U \ll W$. Here, $W$ is the 
band width of the relevant band for pairing. This is in sharp contrast with 
the strong-binding regime, for which $T_\textrm{BCS} \propto U$ is known to 
characterize not the critical SF transition temperature but the pair formation 
one when $U \gg W$, i.e., the naive BCS mean-field theory breaks down.

The standard approach to circumvent this long-known difficulty is to include the 
fluctuations of the order parameter on top of the BCS mean-field in such a way
that the critical SF transition temperature is determined by the universal BKT 
relation through an analogy with the XY model~\cite{b, kt, nk}. Such an analysis 
has recently been carried out for a class of two-band Hamiltonians, including 
the model of interest here~\cite{torma17a}. A compact way to express the 
resultant universal BKT relation is~\cite{torma17a, iskin17, iskin19},
\begin{align}
\label{eqn:bkt}
T_\textrm{BKT} &= \frac{\pi}{8} \sqrt{\det{\boldsymbol{D}}},\\
\label{eqn:conv}
\boldsymbol{D}_{\mu\nu}^\textrm{conv} & = \frac{\Delta^2}{\mathcal{A}} \sum_{s\mathbf{k}} 
\left(
\frac{\mathcal{X}_{s\mathbf{k}}}{E_{s\mathbf{k}}^3} 
- \frac{\mathcal{Y}_{s\mathbf{k}}}{2T E_{s\mathbf{k}}^2}
\right)
\frac{\partial \xi_{s\mathbf{k}}} {\partial k_\mu}
\frac{\partial \xi_{s\mathbf{k}}} {\partial k_\nu}, \\
\boldsymbol{D}_{\mu\nu}^\textrm{geom} & = \frac{2\Delta^2}{\mathcal{A}} \sum_{s \mathbf{k}}
\frac{d_\mathbf{k} \mathcal{X}_{s\mathbf{k}}}{s (\mu - d_\mathbf{k}^0) E_{s\mathbf{k}}} 
\boldsymbol{g}_{\mu\nu}^\mathbf{k},
\label{eqn:geom}
\end{align}
where the tensor
$
\boldsymbol{D} = \boldsymbol{D}^\textrm{conv} + \boldsymbol{D}^\textrm{geom}
$ 
corresponds to the SF phase stiffness, $\mathcal{A}$ is the area of the system, 
and 
$
\mathcal{Y}_{s\mathbf{k}} = \mathrm{sech}^2 [E_{s\mathbf{k}}/(2T)]
$
is a thermal factor. The SF stiffness has two distinct contributions depending 
on the physical processes involved: while the intraband ones are called 
conventional, the interband ones are called geometric since only the latter 
is controlled by the total quantum metric tensor $\boldsymbol{g^\mathbf{k}}$
of the single-particle bands. A compact way to express its elements 
is~\cite{resta11}
\begin{align}
\label{eqn:qmt}
\boldsymbol{g}_{\mu\nu}^\mathbf{k} = \frac{1}{2} 
\frac{\partial \widehat{\mathbf{d}}_\mathbf{k}}{\partial k_\mu}  \cdot 
\frac{\partial \widehat{\mathbf{d}}_\mathbf{k}}{\partial k_\nu},
\end{align}
where 
$
\widehat{\mathbf{d}}_\mathbf{k} = \mathbf{d}_\mathbf{k}/d_\mathbf{k}
$
is a unit vector. Thus, the extended BCS mean-field theory corresponds to 
the self-consistent solutions of Eqs.~(\ref{eqn:op}),~(\ref{eqn:filling}) 
and~(\ref{eqn:bkt}) for $\Delta(T_\textrm{BKT})$, $\mu(T_\textrm{BKT})$ 
and $T_\textrm{BKT}$ for any given set of model parameters 
$t'$, $t''$, $U$, and $F$. In addition to reproducing the well-known BCS 
result for the weak-binding regime where $T_\textrm{BKT} \to T_\textrm{BCS}$ 
from below in the $U \ll W$ limit, this approach provides a reliable 
description of the strong-binding regime where 
$T_\textrm{BKT} \propto W^2/U$ in the $U \gg W$ limit.

\subsection{Strongly-bound molecular pairs}
\label{sec:sbmp}

Since $T_\textrm{BKT} \ll W \ll \Delta$ in the molecular regime, the self-consistency
equations turn out to be analytically tractable, and their closed-form solutions 
provide considerable physical insight into the problem. For instance, 
the order parameter and number equations lead to
$
\Delta = (U/2) \sqrt{F(2-F)}
$
and
$
\mu = -(U/2) (1-F).
$
Similarly, the total SF stiffness reduces to
$
\boldsymbol{D}_{\mu\nu} = \lbrace \Delta^2/[\mathcal{A} (\mu^2+\Delta^2)^{3/2}]\rbrace 
\sum_\mathbf{k} \textrm{Tr} \lbrace
(\partial h_\mathbf{k} / \partial k_\mu)
(\partial h_\mathbf{k} / \partial k_\nu) 
\rbrace
$
with the trace $\textrm{Tr}$ over the sublattice sector, and it is a diagonal 
matrix
$
\boldsymbol{D}_{\mu\nu} = D_0 \delta_{\mu\nu}
$ 
with an isotropic value
$
D_0 = 2\Delta^2 (t^2 + t'^2 + t''^2) /(\mu^2+\Delta^2)^{3/2}.
$
Plugging these expressions into the universal BKT relation, we eventually obtain
$
T_\textrm{BKT} = \pi F (2-F) (t^2 + t'^2 + t''^2)/(2 U),
$
showing that $t'$ and $t''$ increases $T_\textrm{BKT}$ for a given $F$ in 
the $U \gg t$ limit. Furthermore, we may relate $D_0$ to the density 
$\rho_p$ and effective mass $m_p$ of the SF pairs through the relation
$
D_0 = 4\hbar^2 \rho_p / m_p,
$
where
$
\rho_p = F_p/a^2
$
with $F_p$ the filling of SF pairs. To a good approximation, we may identify
$F_ p = F_c/2$ using the well-known expression~\cite{leggett}
\begin{align}
\label{eqn:Fc}
F_c =  \frac{\Delta^2}{2M} \sum_{s \mathbf{k}} 
\frac{\mathcal{X}_{s \mathbf{k}}^2} {E_{s\mathbf{k}}^2}
\end{align}
for the filling of condensed particles. In the strong-binding regime, this 
leads to $F_p \to F_0 = F(2-F)/4$ as the filling of SF pairs whose effective 
mass $m_p = \hbar^2 U/[4a^2(t^2 + t'^2 + t''^2)]$ increases with $U$ but 
decreases with $t'$ and $t''$ in the $U \gg t$ limit. 
In the next section, these analytical expressions are used as a benchmark 
for our numerics, where we explore the solutions of the self-consistency 
equations as a function of $U$.

\begin{figure*}[htbp]
\includegraphics[scale=0.7]{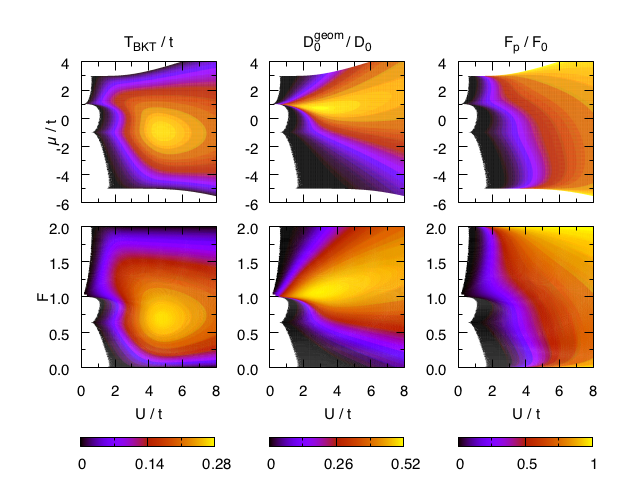}
\caption{(color online)
\label{fig:tp05tpp0}
Left column is the critical SF transition temperature $T_\textrm{BKT}/t$, 
middle column is the relative weight $D_0^\textrm{geom}/D_0$ of the 
geometric SF stiffness, and right column is the faction $F_c/F_0$ of 
condensed particles. Here, the next-nearest-neighbor hoppings are 
$t' = 0.5 t > t'' = 0$, corresponding to the sketch Fig.~\ref{fig:bandsketch}(a).
}
\end{figure*}
\section{Numerical Analysis}
\label{sec:na}

One of our main objectives in this paper is to study how the quantum 
geometry exposes itself in various SF properties through its contribution to 
the SF stiffness given above by Eq.~(\ref{eqn:geom}). This expression suggests 
that a nonzero geometric contribution requires $t' \ne t''$ for a given 
$t \ne 0$, which is simply because the band structure consists effectively 
of a single band when $t' = t''$, and therefore, the interband processes 
necessarily vanish. Thus, for the sake of its conceptual simplicity, 
let us initially set one of the next-nearest-neighbor hopping parameters to 
zero, and consider $t' = 0.5 t > t'' = 0$ as an example. Most important 
features for the corresponding single-particle problem can be extracted 
from the sketch Fig.~\ref{fig:bandsketch}(a), and the self-consistent 
solutions for the many-body problem are presented in Fig.~\ref{fig:tp05tpp0}. 

First of all, having $t' \ne 0$ splits the van Hove singularity of the usual 
square lattice model (for which the singularity lies precisely at $\mu = 0$ 
or half filling $F = 1$ when $t' = t'' = 0$) into two, and produces one 
singularity per band. In Fig.~\ref{fig:tp05tpp0}, these singularities 
are clearly visible at $\mu = \pm t$, and the plus sign 
corresponds also to the energy at which the upper band edges of the lower 
band touch quadratically to the lower band edges of the upper band 
at $F = 1$. In the $U/t \to 0$ limit, we note that the weakly-bound pairs 
first occupy the lower band for $-5t \le \mu < t$ until it is full at $F = 1$, 
and than they occupy the upper band for $t \le \mu < 3t$ until it is full 
at $F = 2$. This figure reveals that it is the competition between these 
singularities that ultimately determines the critical SF transition temperature
$T_\textrm{BKT}$ in the weak-binding regime, i.e., while $T_\textrm{BKT}$ 
is favored by the increase in the single-particle density of states near the van 
Hove singularity of the lower band, it is also enhanced by the geometric 
contribution to the SF stiffness emanating near the band edges/touchings.
Since the band width $2t$ of the upper band is relatively much narrower 
than that $6t$ of the lower band, the geometric contribution plays a more 
decisive role for $F \gtrsim 1$.
For completeness, the fraction $F_p/F_0$ of condensed particles is also 
shown in Fig.~\ref{fig:tp05tpp0}. In comparison to the half filling ($F = 1$) 
where half of the pairs or holes may at most be condensed with $F_0 \to 1/2$ 
in the strong-binding regime, all of the particle (hole) pairs are condensed 
with $F_0 \to F/2$ ($F_0 \to 1-F/2$) in the low particle (hole) filling 
$F \to 0$ ($F \to 2$) limit.

\begin{figure*}[htbp]
\includegraphics[scale=0.7]{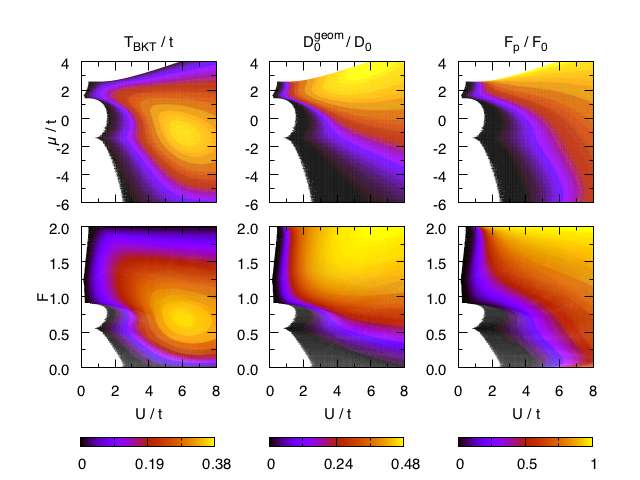}
\caption{(color online)
\label{fig:tp1tpp03}
Same as in Fig.~\ref{fig:tp05tpp0} but for $t' = t$ and $t'' = 0.3t$, 
corresponding to the sketch Fig.~\ref{fig:bandsketch}(b).
}
\end{figure*}

Choosing a different value for $t' < t$ does not change these results 
in any qualitative way, and that the competition between the contributions 
from the van Hove singularity of the lower band and the geometric SF stiffness 
near the band touchings always controls the weak-binding regime.
This also turns out to be the case when both of the next-nearest-neighbor 
hoppings $t' > t'' \ne 0$ are at play. For instance, the self-consistent solutions 
for the $t' = t$ and $t'' = 0.3t$ case are shown in Fig.~\ref{fig:tp1tpp03} as 
an example. Most important features for the single-particle problem 
can be extracted from the sketch Fig.~\ref{fig:bandsketch}(b). In the 
$U/t \to 0$ limit, we find that the weakly-bound pairs first occupy the lower 
band for $-6.6t \le \mu < 1.4t$ until $F \approx 0.92$, and than they 
occupy both bands for $1.4t \le \mu < 2.6t$ until they are full at $F = 2$. 
While the van Hove singularity of the lower band is clearly seen at $\mu = -1.4t$ 
or $F \approx 0.55$, that of the upper band is barely visible around 
$\mu \approx 1.62t$ or $F \approx 1.23$. 
In comparison to Fig.~\ref{fig:tp05tpp0}, since the band width $1.2t$ of 
the upper band is much narrower than that $8t$ of the lower band, and it
lies fully within the energy interval of the latter, the geometric contribution 
plays an even more decisive role for $F \gtrsim 0.9$. In addition, having 
$t'' \ne 0$ also increases the maximal $T_\textrm{BKT}/t$, which is in 
agreement with the analysis given above in Sec.~\ref{sec:sbmp}.

Given our sketches above in Figs.~\ref{fig:bandsketch}(a) and~\ref{fig:bandsketch}(b) 
for the single-particle bands depending on whether the hopping parameters 
are related to each other through $t > t'+t''$ or not, it is possible to reach 
qualitative conclusions for other parameter sets including the flat-band limit 
when $t = t' > t'' = 0$. However, motivated by the growing interest in 
understanding the origins of flat-band superfluidity in various other contexts
discussed above in Sec.~\ref{sec:intro}, next we present an in-depth analysis 
showing how the flat-band superfluidity prevails in our model.

\section{Flat-Band Superfluidity}
\label{sec:fbs}

When $t \le t' + t''$, our sketch Fig.~\ref{fig:bandsketch}(b) implies that 
decreasing the ratio $t''/t$ flattens the upper band with respect to the lower 
one, turning it eventually to an energetically quasi-flat band in the $t''/t \to 0$ 
limit. Here, we first set $t  = t' > t'' = 0$ and analyze the so-called Mielke 
checkerboard lattice model~\cite{mielke91, montambaux18}, and then 
gain more physical insight by studying an effective low-energy continuum 
model for it.

\subsection{Mielke checkerboard lattice model}
\label{sec:mclm}

\begin{figure*}[htbp]
\includegraphics[scale=0.7]{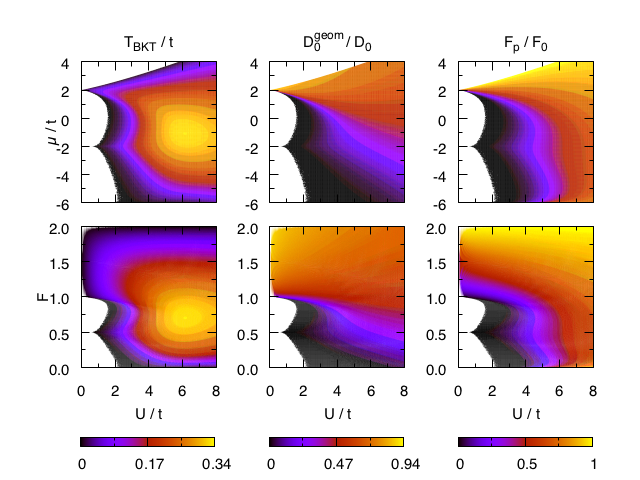}
\caption{(color online)
\label{fig:tp1tpp0}
Same as in Fig.~\ref{fig:tp05tpp0} but for $t' = t > t'' = 0$, corresponding 
to a special limit in the sketch Fig.~\ref{fig:bandsketch}(b), i.e., for 
the so-called Mielke checkerboard lattice model discussed in 
Sec.~\ref{sec:mclm}. 
}
\end{figure*}

As discussed in Sec.~\ref{sec:mfh}, the band structure in the Mielke checkerboard 
lattice is simply determined by
\begin{align}
\label{eqn:mcup}
\varepsilon_{+, \mathbf{k}} &= 2t, \\
\label{eqn:mcdo}
\varepsilon_{-, \mathbf{k}} &= -2t - 4t \cos(k_xa) \cos(k_ya),
\end{align}
for which the van Hove singularity of the lower band is precisely at 
$\varepsilon = -2t$ or $F = 0.5$. This is in such a way that the lower band lies 
within the energy interval $-6t \le \varepsilon_{-, \mathbf{k}} < 2t$, and its upper 
band edges touch to the flat band at the four corners of the first Brillouin zone.
All of these features are clearly visible in the self-consistent solutions for the 
many-body problem that are presented in Fig.~\ref{fig:tp1tpp0}.
In the $U/t \to 0$ limit, we find that the weakly-bound pairs first occupy 
the lower band for $-6t \le \mu < 2t$ until $F = 1$, and than they occupy 
the upper band at $\mu = 2t$ until it is full at $F = 2$. Since the upper band
is entirely flat, the geometric contribution dominates the parameter space 
for $F \gtrsim 1$ in especially the weak-binding regime. 

Thanks to its analytical tractability, next we focus only on the $\mu = 2t$ case, 
and reveal the origins of flat-band superfluidity as a function of $U/t$.
For instance, when $\mu$ coincides with an isolated flat band 
at $U = 0$~\cite{iskin17f}, the ground state (i.e., at $T = 0$ the zero 
temperature) is determined by
$
\Delta = (U/2)\sqrt{f(1-f)}
$ 
and
$
\mu = 2t + (U/2)(f-1/2)
$ 
as soon as $U/t \ne 0$. 
Here, $0 \le f = F - 1 \le 1$ is the filling of the 
isolated flat band for which $F_c = f(1-f)$ gives the filling of condensed 
particles. Thus, setting $\mu = 2t$ in the $U/t \to 0$ limit, we find $f=1/2$, 
$\Delta = U/4$, $F = 3/2$ and $F_c = 1/4$, which are in perfect 
agreement with our numerics. Furthermore, our numerical calculations 
for the conventional contribution Eq.~(\ref{eqn:conv}) to the SF stiffness
$
\boldsymbol{D}_{\mu\nu}^\textrm{conv} = (\Delta^2/\mathcal{A}) 
\sum_{\mathbf{k}} 
(\partial \xi_{-,\mathbf{k}} / \partial k_\mu)
(\partial \xi_{-,\mathbf{k}} / \partial k_\nu)/E_{-,\mathbf{k}}^3
$
in the ground state fit perfectly well with
$
D_0^\textrm{conv}/(2F_c) = U/(4\pi)
$
when $\Delta/t \to 0$ in the $U/t \to 0$ limit. Similarly, the geometric
contribution Eq.~(\ref{eqn:geom}) to the SF stiffness
$
\boldsymbol{D}_{\mu\nu}^\textrm{geom}  
= (2\Delta/\mathcal{A}) \sum_{\mathbf{k}}
\boldsymbol{g}_{\mu\nu}^\mathbf{k} (1 - \Delta/E_{-,\mathbf{k}})
$
in the ground state fit extremely well (i.e., up to the machine precision) with
$
D_0^\textrm{geom}/(2F_c) = U [\ln(64t/U) - 1]/(4\pi)
$
when $\Delta/t \to 0$ in the $U/t \to 0$ limit. 
In comparison, one can also calculate that
$
D_0^\textrm{geom}/(2F_c) = D_0^\textrm{conv}/(2F_c) = 4t^2/U
$
when $\Delta/t \gg 1$ in the $U/t \gg 1$ limit, which is in agreement 
with the analysis given above in Sec.~\ref{sec:sbmp} where 
$F_c = F(2-F)/2 \to 0.5$ in the $F \to 1$ limit when $\mu = 2t$. 

\begin{figure}[htbp]
\includegraphics[scale=0.6]{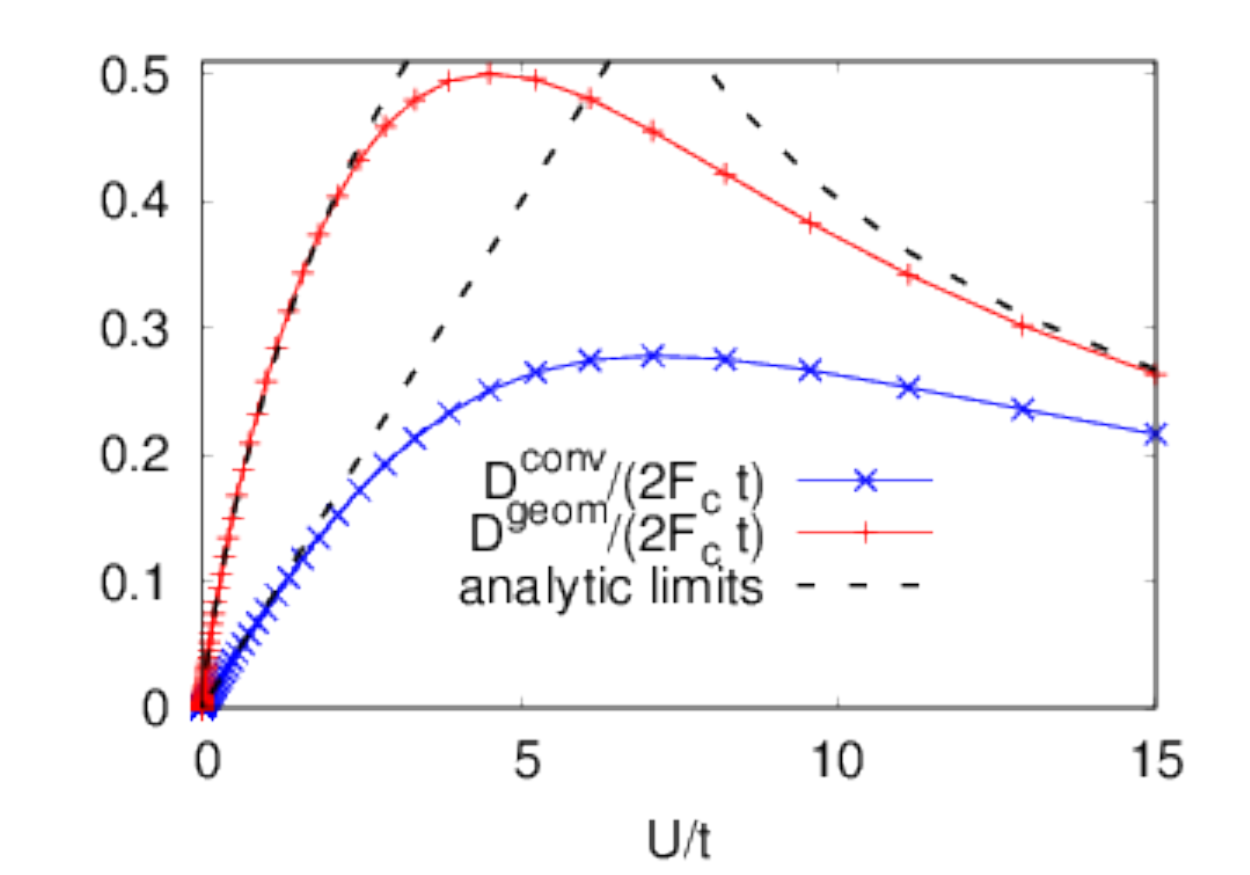}
\caption{(color online)
\label{fig:mu2}
Conventional and geometric contributions to the SF stiffness for the 
Mielke checkerboard lattice model. Here, we set $t' = t > t'' = 0$ and 
$\mu = 2t$, and evaluate the contributions self-consistently at 
$T = T_\textrm{BKT}$.
}
\end{figure}

We also observe that the very same analytical expressions, i.e., 
$
D_0^\textrm{conv}/(2F_c) = U/(4\pi)
$
for the conventional contribution and
$
D_0^\textrm{geom}/(2F_c) = U [\ln(64t/U) - 1]/(4\pi)
$
for the geometric contribution, fit extremely well with the numerical 
results at $T = T_\textrm{BKT}$ in the $U/t \to 0$ limit. Here, $F_c$ also 
needs to be evaluated at $T_\textrm{BKT}$, for which Eq.~(\ref{eqn:Fc})
leads to $F_c \to 0$ when $\Delta/t \to 0$ in the $U/t \to 0$ limit.
This observation is illustrated in Fig.~\ref{fig:mu2}, where we plot the 
self-consistent numerical solutions together with the analytical fits. 
We note that the $U$ dependence of the total SF stiffness
$
D_0/(2F_c) = U \ln(64t/U)
$
is very different from those $D_0/(2F_c) \propto U $ of the isolated
flat bands~\cite{torma15, torma16, torma17a}, which is one of the 
direct consequences of the band touchings as discussed below
in Sec.~\ref{sec:elcm}.

The main physical reason behind the success of the very same fit at 
both $T = 0$ and $T = T_\textrm{BKT}$ has to do with the effective mass 
$m_p$ of the strongly-bound molecular pairs. This is because, in 
accordance with the analysis given above in Sec.~\ref{sec:sbmp}, one may 
define the inverse of the effective pair-mass tensor $\boldsymbol{m_p}$ 
through plugging $\rho_p = F_c/(2 a^2)$ for the density of SF pairs in the 
relation
\begin{align}
\label{eqn:mp}
\boldsymbol{D} = 4\hbar^2 \rho_p \boldsymbol{m_p^{-1}}.
\end{align}
Thus, our analytical fits suggest that the effective mass of the pairs diverges as
$
m_p = 4\pi \hbar^2 / [U a^2 \ln(64t/U)]
$
when $\mu = 2t$ in the $U/t \to 0$ limit, which is directly caused by the 
diverging effective band mass of a single particle in a flat band. 
Furthermore, by separating the $\boldsymbol{D}$ and $\boldsymbol{m_p^{-1}}$ 
tensors into their conventional and geometric contributions, i.e., 
$
(\boldsymbol{m_p^{-1}})_{\mu\nu} = 
(\boldsymbol{m_p^{-1}})_{\mu\nu}^\textrm{conv} +
(\boldsymbol{m_p^{-1}})_{\mu\nu}^\textrm{geom},
$
and using Eq.~(\ref{eqn:geom}) along with the ground state parameters, 
we may also identify
\begin{align}
\label{eqn:mpg}
(\boldsymbol{m_p^{-1}})_{\mu\nu}^\textrm{geom}
= \frac{U}{\hbar^2 M} \sum_{\mathbf{k}} \boldsymbol{g}_{\mu\nu}^\mathbf{k} 
\left(1 - \frac{1}{\sqrt{1 + \frac{16 \xi_{-,\mathbf{k}}^2}{U^2}}} \right)
\end{align}
in the $U/t \to 0$ limit. This expression is almost identical to a very recent 
result~\cite{torma19} where the inverse mass tensor of the two-body problem 
in an isolated flat band is reported as
$
(\boldsymbol{m_p^{-1}})_{\mu\nu}^\textrm{geom}  
= [U/(\hbar^2 M)] \sum_{\mathbf{k}} \boldsymbol{g}_{\mu\nu}^\mathbf{k}.
$
Since the Mielke flat band is not entirely isolated from the lower band, 
there is an extra term in Eq.~(\ref{eqn:mpg}) that cancels precisely those 
band touchings from the sum, i.e., when $\xi_{-,\mathbf{k}} \to 0$ in the 
$U/t \to 0$ limit. To show that the sum
$
\sum_{\mathbf{k}} \boldsymbol{g}_{\mu\nu}^\mathbf{k}
$
by itself is infrared divergent for a flat band that is in touch with a dispersive 
one, next we construct a low-energy continuum model for the Mielke 
checkerboard lattice.

\subsection{Effective low-energy continuum model}
\label{sec:elcm}

\begin{figure}[htbp]
\includegraphics[scale=0.6]{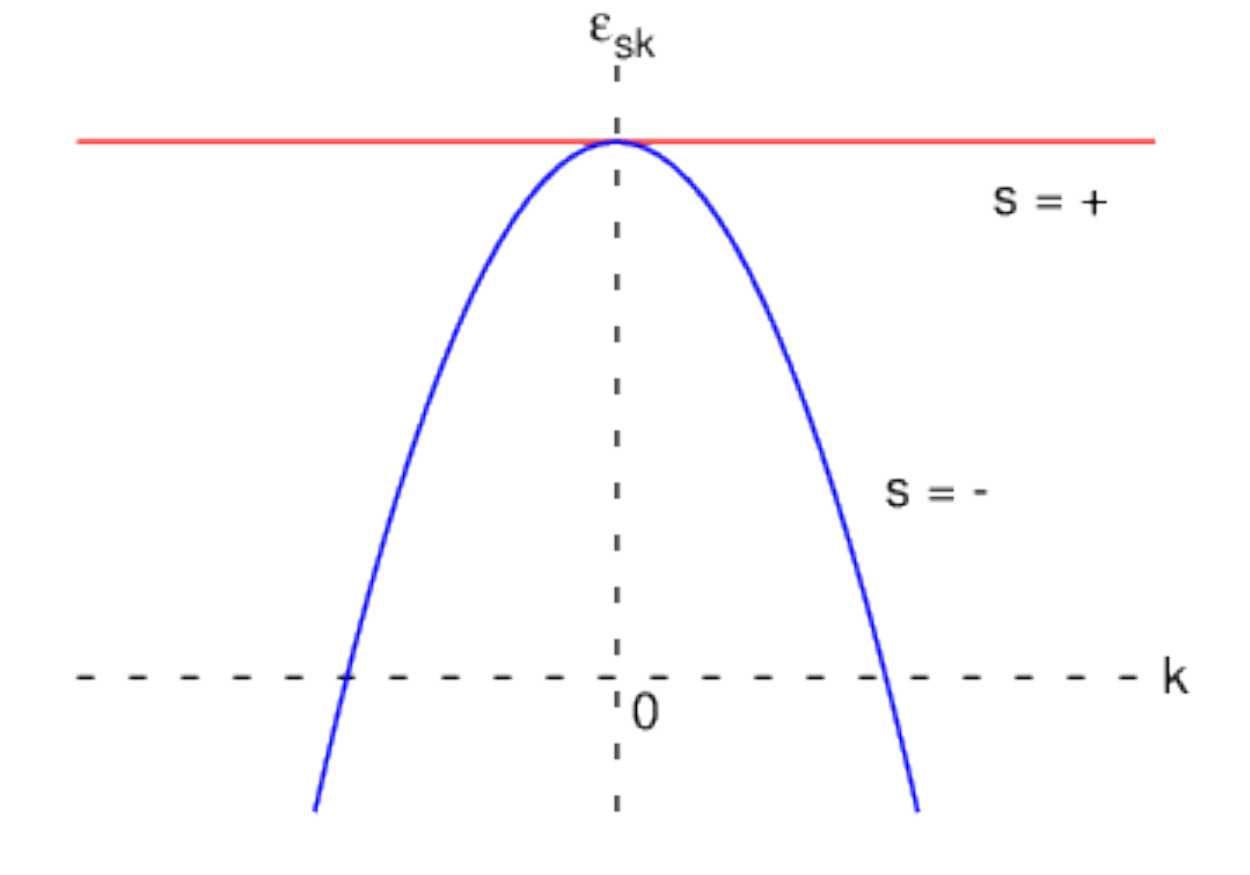}
\caption{(color online)
\label{fig:bands}
Band structure of the effective low-energy continuum model discussed in
Sec.~\ref{sec:elcm}.
}
\end{figure}

For this purpose, we expand the single-particle Hamiltonian density 
$h_\mathbf{k}$ given above by Eqs.~(\ref{eqn:hk})-(\ref{eqn:dkz}) near the 
band touchings, and arrive at its low-energy description where
\begin{align}
\label{eqn:lowd0k}
d_\mathbf{k}^0 &= 2t - t (k_x^2+k_y^2)a^2, \\
\label{eqn:lowdxk}
d_\mathbf{k}^x &= - t (k_x^2 - k_y^2)a^2, \\
\label{eqn:lowdzk}
d_\mathbf{k}^z &= - 2t k_x k_ya^2.
\end{align}
Then, the band structure is simply determined by
$
\varepsilon_{+, \mathbf{k}} = 2t
$
for the upper band and
$
\varepsilon_{-, \mathbf{k}} = 2t - 2 t k^2 a^2
$
for the lower band, exhibiting a quadratic touching point at 
$\mathbf{k} = \mathbf{0}$ as sketched in Fig.~\ref{fig:bands}. 
In addition, given that the band geometry is also characterized by a 
much simpler quantum metric tensor, i.e., Eq.~(\ref{eqn:qmt}) reduces to
\begin{align}
\label{eqn:lowqm}
\boldsymbol{g}_{\mu\nu}^\mathbf{k} = 2\frac{k^2 \delta_{\mu\nu} - k_\mu k_\nu}{k^4},
\end{align}
one can gain further physical insight by studying this continuum model 
at $\mu = 2t$ in the $U/t \to 0$ limit.

In perfect agreement with Sec.~\ref{sec:mclm}, the ground state is again
determined by $\Delta = U/4$ and $F_c = 1/4$, and we find
$
D_0^\textrm{conv} = \Delta/(2\pi)
$
for the conventional contribution. Furthermore, we find that the
infrared divergence of the sum
$
\sum_\mathbf{k} \boldsymbol{g}_{\mu\nu}^\mathbf{k} 
= [\mathcal{A}/(2\pi)] \ln(k_{\max}/k_{\min}) \delta_{\mu\nu}
$
is precisely cancelled by the infrared divergence of the sum
$
\Delta \sum_\mathbf{k} \boldsymbol{g}_{\mu\nu}^\mathbf{k}/E_{-,\mathbf{k}} 
= - [\mathcal{A}/(4\pi)] \ln(t k_{\min}^2 a^2/\Delta) \delta_{\mu\nu},
$
assuming
$
2 t k_{\min}^2 a^2 \ll \Delta \ll 2 t k_{\max}^2 a^2
$ 
in the $U/t \to 0$ limit.
Here, $W = 2 t k_{\max}^2 a^2$ is the band width of the lower band. 
Note that the analytical fit given above in Sec.~\ref{sec:mclm} implies that
$
k_{\max} a \simeq 4/\sqrt{e} \approx 2.426.
$
However, using the relation 
$
\sum_{s \mathbf{k}} 1 = M
$
for the number of lattice sites, we find
$
k_{\max} a = \sqrt{2\pi} \approx 2.507
$ 
and $W = 4\pi t$, leading eventually to
$
D_0^\textrm{geom} = [\Delta/(2\pi)] \ln[W/(2 \Delta)].
$
Thus, the low-energy model explains most of our findings in 
Sec.~\ref{sec:mclm}.

\section{Conclusions}
\label{sec:conc}

To summarize, we exposed the quantum-geometric origin of flat-band 
superfluidity in the Mielke checkerboard lattice whose two-band band 
structure consists of an energetically flat band that is in touch with a 
quadratically dispersive one, i.e., a non-isolated flat band. For instance, 
in the weak-binding regime of arbitrarily low $U/t$, we found that 
the inverse effective mass tensor $\boldsymbol{m_p^{-1}}$ of the 
Cooper pairs is determined entirely by a $\mathbf{k}$-space 
sum over the so-called quantum metric tensor of 
the single-particle bands, leading to
$
m_p = 4\pi \hbar^2 / [U a^2 \ln(64t/U)]
$
in the $U/t \to 0$ limit. Since the effective band mass of a non-interacting 
particle is infinite in a flat band, this particular result illuminates the physical
mechanism behind how the mass of the SF carriers becomes finite with 
a finite interaction, i.e., how the quantum geometry is responsible for 
$m_p \ne \infty$ through interband processes as soon as $U/t \ne 0$. 
When $U/t$ increases from $0$, we also found that the geometric 
interband contribution gradually gives way to the conventional intraband 
one, eventually playing an equally important role in the strong-binding 
regime when $U/t \gg 1$.

Furthermore, given that $m_p$ plays direct roles in a variety of SF properties
that are thoroughly discussed in this paper (i.e., the SF stiffness and 
critical SF transition temperature) but not limited to them 
(e.g., the sound velocity), this result also illuminates how the mean-field 
BCS correlations prevail in a non-isolated flat band. Curiously enough, 
such revelations of a fundamental connection between a physical 
observable and the underlying quantum geometry turn out to be quite 
rare in nature~\cite{resta11}, making their cold-atom realization a gold mine 
for fundamental physics. We hope that our work motivates further research 
along these lines.

\begin{acknowledgments}
This work is supported by the funding from 
T{\"U}B{\.I}TAK Grant No. 1001-118F359.
\end{acknowledgments}

\end{document}